\documentstyle[aps,manuscript,graphicx]{revtex}

\begin{document}

\title{Charmonium suppression by gluon bremsstrahlung in p-A and A-B collisions}

\author{J. H\"ufner, Y.B. He}

\address{Institut f\"ur Theoretische Physik, Universit\"at Heidelberg, }

\address{Philosophenweg 19, D-69120 Heidelberg, Germany}

\author{B.Z. Kopeliovich}

\address{Max-Planck Institut f\"ur Kernphysik, Postfach 103980, 69029 Heidelberg, Germany}

\address{Joint Institute for Nuclear Research, Dubna, 141980 Moscow Region, Russia}

\maketitle
\begin{abstract}
Prompt gluons are an additional source for charmonium suppression in nuclear
collisions, in particular for nucleus-nucleus collisions. These gluons are radiated
as bremsstrahlung in N-N collisions and interact inelastically with the charmonium
states while the nuclei still overlap. The spectra and mean number \( \langle n_{g}\rangle  \)
of the prompt gluons are calculated perturbatively and the inelastic cross section
\( \sigma _{\mathrm{abs}}^{\Psi g} \) is estimated. The integrated cross sections
\( \sigma (A\, B\longrightarrow J/\psi \, X) \) for p-A and A-B collisions
and the dependence on transverse energy for S-U and Pb-Pb can be described quantitatively
with some adjustment of one parameter \( \langle n_{g}\rangle \, \sigma _{\mathrm{abs}}^{\Psi g} \). 
\end{abstract}

\section{Introduction}

The original proposal \cite{Mat86} of using charmonium production in heavy-ion
collisions as a signal of QGP formation has triggered a series of experimental
measurements at the CERN SPS in proton-nucleus and nucleus-nucleus reactions(c.f.
the reviews\cite{Rog99,Ger99}). One remarkable observation out of these experiments\cite{NA5198,NA3898,NA3899}
is: when one scans through data from the lighter to the heavier interacting
nuclei, one sees that the \( J/\psi  \) suppression in p-p up to central S-U
reactions is consistent with a picture of pure final-state absorption on nucleons,
while the data for central Pb-Pb collisions deviate from this systematics and
are therefore termed ``anomalous''.

Among the many theoretical concepts we mention the following three mechanisms
for \( \Psi  \) suppression which have been proposed (here \( \Psi  \) stands
for \( J/\psi , \) \( \chi _{c} \) , \( \psi ^{\prime } \)): 

\begin{enumerate}
\item Inelastic collisions of a charmonium premeson \( c\bar{c} \) by the nucleons
of projectile and target. The inelastic cross section \( \sigma _{\mathrm{abs}}^{c\bar{c}\, \Psi }(\tau ) \)
is a function of the time \( \tau  \) (measured in the rest system of the premeson
until the asymptotic meson is formed)\cite{Bro88,Kop91,Aic99,He99}. In most
calculations an absorption cross section with adjusted value of \( \sigma _{\mathrm{abs}}=6 \)
mb, independent of time, is used both for the \( J/\psi  \) and \( \psi ^{\prime } \)
channel. While this approach reproduces the suppression in p-A collisions and
A-B collisions with small projectiles, central Pb-Pb collisions show stronger
suppressions. 
\item Charmonium dissociation in the quark-gluon plasma phase of the reaction. The
interaction of charmonium with the partons, in particular the gluons leads to
suppression. The different binding energies of the states \( J/\psi , \) \( \chi _{c} \),
and \( \psi ^{\prime } \) play a crucial role and may lead to discontinuous
behaviour in the \( E_{T} \) dependence of the suppression\cite{Bla96,Won97,Kha97,Vog98}. 
\item In the final state of a nucleus-nucleus collision many hadrons emerge from the
reaction volume together with the charmonium. The interaction with these comoving
hadrons is an additional source of \( \Psi  \) suppression\cite{Gav97,Cap97,Arm99,Cas97}.
In approaches of this kind, the inelastic cross section \( \sigma _{\mathrm{co}}^{\Psi } \)
of the \( \Psi  \) with the comovers is the crucial adjusted parameter for
which no reliable calculations are available at present. 
\end{enumerate}
In this situation we propose still another mechanism, suppression by prompt
gluons. These gluons are produced as bremsstrahlung in the N-N collisions in
the early phase of the reaction. Those gluons, whose production time is sufficiently
short, will interact inelastically with the charmonium in addition to and at
the same time as the \( \Psi  \)-N interactions. In another language: The gluons
produced in a N-N collision, \( N+N\longrightarrow g+N+N \) may be seen together
with N as a wounded nucleon \( N^{*}=N+g \) . In this language a \( \Psi  \)
will be suppressed by collisions with nucleons \( N \) in their ground state
and by wounded nucleons, \( N^{*}. \) Whatever picture one prefers, the prompt
gluons will be an additional source of suppression. Their number and their momentum
distribution can be calculated perturbatively. Uncertainties remain for the
inelastic cross section \( \sigma _{\mathrm{abs}}^{\Psi g} \).

In this paper we expand on the preliminary calculations which were presented
in Ref. \cite{hue98}. While the general idea is the same, calculations of the
\( \Psi  \) suppression by nucleons and prompt gluons only, several improvements
have been introduced and more observables have been calculated:

(i) A closed expression is derived within a classical multiple scattering model
with straight line trajectories.

(ii) Formation time effects are included for the produced premeson and production
time effects for the gluon by introducing a time dependence for the inelastic
cross sections \( \sigma _{\mathrm{abs}}^{\Psi N}(t) \), \( \sigma _{\mathrm{abs}}^{\Psi g}(t) \)
and for the mean numbers \( \langle n_{g}\rangle (t) \) of prompt gluons. 

(ii) We also calculate charmonium suppression in its dependence on transverse
energy;

(iv) We predict suppressions for reactions with inverse kinematics.

\section{Formalism}

\begin{figure}[tbh]
\begin{center}
\includegraphics[height=10cm]{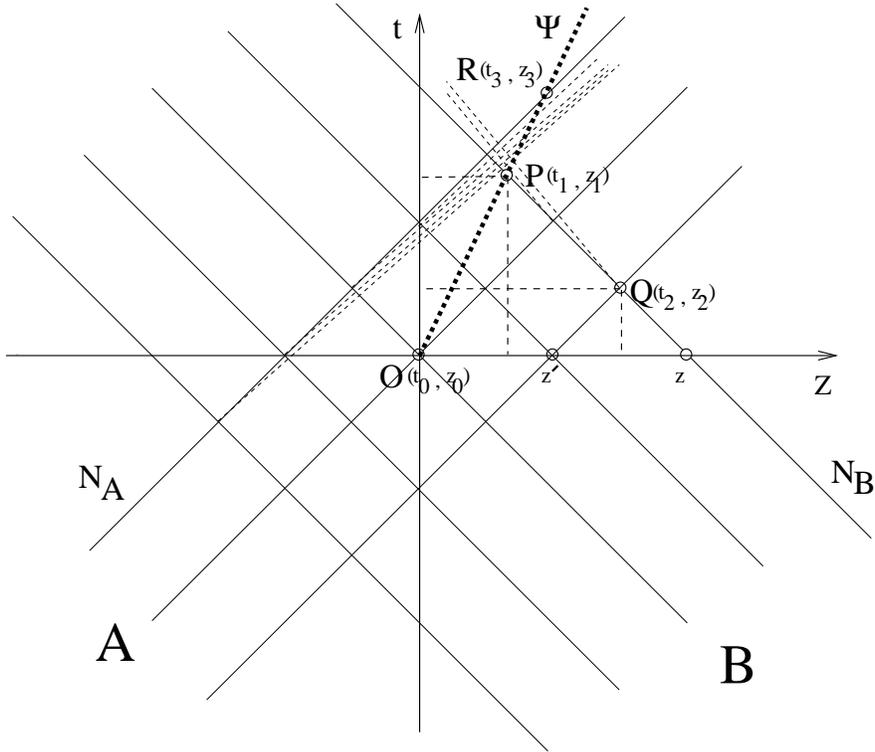}
\caption{The collision between a projectile nucleus A with a 
target nucleus B in the
two dimensional time-longitudinal coordinate representation in the NN c.m. system.
Nucleons are denoted by solid lines, the premeson 
produced at $O(t_{0,}z_{0})$
is denoted by heavy dots, while the bremsstrahl gluons are drawn by light 
dashes.\label{ABfig}}
\end{center}
\end{figure}

We follow the original idea of Ref.~\cite{hue98} and derive a closed expressions
for charmonium suppression caused by nucleons and prompt gluons. The derivation
is based on Fig.\ref{ABfig}, which displays a collision between a projectile
nucleus A and a target nucleus B in the two-dimensional \( t-z \) (time-longitudinal
coordinate) plane and in the NN c.m.s.. Since projectile and target nucleons
have high energies (for the 200~GeV/A energy at SPS, projectile and target are
characterized by a Lorentz-factor \( \gamma =10 \) ), the nucleons move nearly
on the light-cone. Assuming zero production time for the premeson \cite{He99},
a premeson is produced at point \( O(t_{0},z_{0}) \) and moves with a velocity
\( v_{cm} \) in the c.m.s.. During its motion along a straight line trajectory
its internal structure develops in time towards the asymptotically observed
\( \Psi . \) This evolution, called formation, is characterized by a certain
formation time and manifests itself in a time dependent absorption cross section
\( \sigma _{\mathrm{abs}}^{\Psi N}(t). \) It interacts at point \( P(t_{1},z_{1}) \)
with a nucleon from the target B. Before this encounter, this nucleon has experienced
a collision at \( Q(t_{2},z_{2}) \) with another nucleon from the projectile,
leading to the radiation of a prompt gluon (dashed line) which also interacts
with the \( \Psi  \) at \( P(t_{1},z_{1}). \) The time difference \( \Delta t_{\Psi }=t_{1}-t_{0} \)
(measured in the NN c.m.s.) has to be put into relation to the formation time
of the charmonium. Similarly the time difference \( \Delta t_{g}=t_{1}-t_{2} \)
has to be compared to the time it takes to produce a gluon (coherence time),
which manifests itself in a time dependent mean number of gluons \( \langle n_{g}(\Delta t_{g})\rangle  \).
At the point \( R(t_{3},z_{3}) \) the \( \Psi  \) interacts with a projectile
nucleon \( N_{A} \) and its comoving prompt gluons. 

First we treat the interaction of the premeson with target nucleon. The velocity
of nucleons is taken as 1 (in unit of \( c \)). After created at point \( O(t_{0,}z_{0}) \)
but before colliding with the target nucleon \( N_{B} \) at point \( P(t_{1},z_{1}) \),
the premeson \( \Psi  \) has lived for a time interval \( \Delta t_{\Psi }=t_{1}-t_{0}
=(z-z_{0})/(1+v_{cm}) \).
Therefore, the contribution to \( \Psi  \) suppression from the target nucleons
can be taken into account by an attenuation factor \( e^{-X} \) with

\begin{equation}
\label{CNeq}
X=\int _{z_{0}}^{\infty }dz\, \gamma \, \rho _{B}(z)\, 
\sigma _{\mathrm{abs}}^{\Psi N}(\Delta t_{\Psi })\, \, ,
\end{equation}
 where we have suppressed the impact parameter dependence of \( \rho _{B} \),
the density of the target nucleus B. The time dependent inelastic cross section
\( \sigma _{\mathrm{abs}}^{\Psi N}(\Delta t_{\Psi }) \) accounts for the formation
of the charmonium state (see below). 

We treat the interaction with prompt gluons in the following way: The target
nucleon \( N_{B} \) collides with a projectile nucleon at point \( Q(t_{2},z_{2}) \)
and a gluon is radiated. This radiation process is not instantaneous, but is
governed by a characteristic time, the production or coherence time (see below).
The gluon radiated at \( Q(t_{2},z_{2}) \) interacts with the \( \Psi  \)
at the point \( P(t_{1},z_{1}) \) after a time interval,

\[
\Delta t_{g}=\Delta t_{\Psi }-\frac{z-z^{\prime }}{2}\, .\]
 The contribution to \( \Psi  \) suppression from gluons is therefore given
by an attenuation factor \( e^{-Y} \) with

\begin{equation}
\label{Cgeq}
Y=\int _{z_{0}}^{\infty }dz\, \gamma \, \rho _{B}(z)\, 
\sigma _{\mathrm{abs}}^{\Psi g}(\Delta t_{\Psi })\, \int _{-\infty }^{+\infty }dz^{\prime }\, \Theta (\Delta t_{g})\, \gamma \, \rho _{A}(z^{\prime })\, \sigma _{\mathrm{in}}^{NN}\, \langle n_{g}(\Delta t_{g})\rangle \, \, ,
\end{equation}
 where \( \sigma _{\mathrm{in}}^{NN} \) is the inelastic N-N cross section,
\( \langle n_{g}(\Delta t_{g})\rangle  \) the mean number of gluons radiated
in a N-N interaction (to be discussed below), and the \( \Theta  \)-function
ensures that only gluons, which are created before the interaction with the
\( \Psi  \), are taken into account.

The suppression of the \( \Psi  \) via the interaction with projectile nucleons
and their accompanying gluons can be derived by appropriate variable exchanges.
Combining these results together one can write down the \( \Psi  \) suppression
factor in \( A \) -\( B \) collisions as:

\begin{equation}
\label{SABEq}
S_{AB}^{\Psi }=\frac{1}{AB}\int d^{2}b\int d^{2}s\int _{-\infty }^{+\infty }dz_{A}\, \gamma \, \int _{-\infty }^{+\infty }dz_{B}\, \gamma \, \rho _{A}(z_{A},\vec{s})\, \rho _{B}(z_{B},\vec{b}-\vec{s})\, \exp [-I_{1}-I_{2}]\, ,
\end{equation}
 where \( I_{1} \) is given by

\begin{eqnarray}
I_{1} & = & \int _{z_{A}}^{+\infty }dz\, \gamma \, \rho _{A}(z,\vec{s})\, \sigma _{\mathrm{abs}}^{\Psi N}(\Delta t_{\Psi })\nonumber \\
 &  & \times \left[ 1+\frac{\sigma _{\mathrm{abs}}^{\Psi g}}{\sigma _{\mathrm{abs}}^{\Psi N}}\int _{-\infty }^{+\infty }dz^{\prime }\, \gamma \, \Theta (\Delta t_{g})\, \rho _{B}(z^{\prime },\vec{b}-\vec{s})\, \sigma _{\mathrm{in}}^{NN}\, \langle n_{g}(\Delta t_{g})\rangle \right] \, \, ,\label{I1Eq} 
\end{eqnarray}
 and \( I_{2} \) is derived from \( I_{1} \) by the substitutions: \( z_{A}\rightarrow z_{B,} \)
\( \rho _{A}(z,\vec{s})\rightarrow \rho _{B}(z,\vec{b}-\vec{s}), \) \( \rho _{B}(z^{\prime },\vec{b}-\vec{s})\rightarrow \rho _{A}(z^{\prime },\vec{s}), \)
\( v_{cm}\rightarrow -v_{cm} \) (hidden in \( \Delta t_{\Psi } \)).

The first and second terms in the exponential in Eq. (\ref{SABEq}) correspond
to contributions from the nuclei A and B, respectively. In the absence of prompt
gluons, \( \langle n_{g}\rangle =0, \) and neglecting the dependence on \( \Delta t_{\Psi } \),
one has the usual expression for charmonium suppression by nucleons. The \( E_{T} \)
dependence of the charmonium suppression is calculated from the impact parameter
dependent suppression function
\begin{equation}
\label{SAB_bEq}
S_{AB}^{\Psi }(\vec{b})=\frac{\int d^{2}s\int _{-\infty }^{+\infty }dz_{A}\, \gamma \, \int _{-\infty }^{+\infty }dz_{B}\, \gamma \, \rho _{A}(z_{A},\vec{s})\, \rho _{B}(z_{B},\vec{b}-\vec{s})\, \exp [-I_{1}-I_{2}]}{\int d^{2}s\int _{-\infty }^{+\infty }dz_{A}\, \gamma \, \int _{-\infty }^{+\infty }dz_{B}\, \gamma \, \rho _{A}(z_{A},\vec{s})\, \rho _{B}(z_{B},\vec{b}-\vec{s})}\, \, .
\end{equation}
 The suppression for a given value \( E_{T} \) is then calculated from
\[
S_{AB}^{\Psi }(E_{T})=\int d^{2}b\, P(\vec{b};E_{T})\, S_{AB}^{\Psi }(\vec{b})\, \, ,\]
 where the distribution \( d^{2}b\, P(\vec{b};E_{T}) \) gives the probability
that a given impact parameter contributes to a particular value of \( E_{T}. \)
We use the following simplification: The NA50/51 group has made simulations
and give the mean value \( \langle b\rangle (E_{T}) \) for each values of \( E_{T} \)
where data exist. We have inserted these values into Eq.~(\ref{SAB_bEq}).

Proton-nucleus collisions can be treated as a special case of Eq. (\ref{SABEq}).

For the time dependent effective cross section \( \sigma _{\mathrm{abs}}^{\Psi N}(\Delta t_{\Psi }) \)
between the charmonium state \( \Psi  \) and a nucleon the expression of the
two-channel model for the evolution of a \( c\bar{c} \) pair is used\cite{hue96,He99}:

\begin{equation}
\label{sigmaeffEq}
\sigma _{\mathrm{abs}}^{\Psi N}(\tau )=\sigma ^{\Psi N}_{\mathrm{in}}+(\sigma ^{\mathrm{pre}}_{\mathrm{in}}-\sigma ^{\Psi N}_{\mathrm{in}})\cos (\Delta M\, \tau )]
\end{equation}
 where \( \sigma ^{\Psi N}_{\mathrm{in}} \) is the asymptotic (\( \tau \rightarrow \infty  \))
\( \Psi  \)-\( N \) cross section, \( \sigma ^{\mathrm{pre}}_{\mathrm{in}} \)
the initial (\( \tau \rightarrow 0 \)) premeson-N cross section, and \( \Delta M=M_{\psi ^{\prime }}-M_{J/\psi } \).
We use values \( \sigma _{\mathrm{in}}^{J/\psi N}=6.7 \) mb (which includes
the contributions of \( \chi _{c} \) and \( \psi ^{\prime } \) ) and \( \sigma _{\mathrm{in}}^{\psi ^{\prime }N}=12 \)
mb. For \( \tau \rightarrow 0 \) we have the absorption cross section of the
premeson, \( \sigma ^{\mathrm{pre}}_{\mathrm{in}} \) , which we take the same
in the \( J/\psi  \) and \( \psi ^{\prime } \) channels to be 3 mb\cite{He99}.

The mean number \( \langle n_{g}\rangle  \) of gluons radiated in a \( N \)
-\( N \) collision in the direction of one of the two nucleons has been calculated
in \cite{hue98}:

\begin{equation}
\label{ngeq}
\langle n_{g}(t_{g})\rangle =\frac{3}{\sigma _{in}^{NN}}\int _{k_{min}^{2}}^{\infty }dk^{2}\int _{\alpha _{min}}^{1}d\alpha \, \frac{d\sigma (qN\rightarrow gX)}{d\alpha dk^{2}}\, \left[ 1-\exp \left( -\frac{t_{g}}{t_{c}^{g}}\right) \right] \, \, ,
\end{equation}
where\cite{Kop98}
\[
\frac{d\sigma (qN\rightarrow gX)}{d\alpha dk^{2}}=\frac{3\alpha _{s}(k^{2})\, C}{\pi }\frac{2m_{q}^{2}\alpha ^{4}k^{2}+[1+(1-\alpha )^{2}](k^{4}+\alpha ^{4}m_{q}^{4})}{(k^{2}+\alpha ^{2}m_{q}^{2})^{4}}\left[ \alpha +\frac{9}{4}\frac{1-\alpha }{\alpha }\right] \, \, .\]
 Here \( \alpha _{s}(k^{2}) \) is the QCD running coupling constant; \( C \)
is the factor for the dipole approximation for the cross section of a \( \bar{q}q \)
pair with a nucleon, \( \sigma ^{\bar{q}q}(r_{T})\approx C\, r_{T}^{2}, \)
where \( r_{T} \) is the \( \bar{q}q \) transverse separation \cite{Zam81,pov87,Kop98}.
\( C\approx 3 \) is the pQCD prediction. The coherence time is given by

\[
t_{c}^{g}=\frac{2E_{q}\alpha (1-\alpha )}{\alpha ^{2}m_{q}^{2}+k^{2}}\, \, ,\]
 and the lower limit of the \( \alpha  \)-integration

\[
\alpha _{\mathrm{min}}=\left\{ \begin{array}{ll}
(\omega _{\mathrm{min}}+\sqrt{\omega _{\mathrm{min}}^{2}-k^{2}})/(2E_{q}) & \, \, \, \, \, \, \mathrm{if}\, \, k<\omega _{\mathrm{min}}\, ,\\
k/(2E_{q}) & \, \, \, \, \, \, \mathrm{otherwise}.
\end{array}\right. \]
 Here \( \alpha  \) is the fraction of the quark light-cone momentum and \( k \)
the transverse momentum carried by the gluon. Compared to Ref. \cite{hue98}
we have introduced a time dependence into the mean number of gluons by inserting
the square bracket into the integral Eq. (\ref{ngeq}). While the limits \( t\rightarrow 0 \)
and \( t\rightarrow \infty  \) and the characteristic time \( t_{c}^{g} \)
for the rise are certainly correct, the explicit form of the exponential is
arbitrary. In our calculation we use \( k\geq 0.5 \) GeV/\( c \) and thus
restrict ourselves to (semi-)hard gluons. Their energy \( \omega \geq \omega _{\mathrm{min}} \)
has to be chosen in relation to the type of charmonium. For example, the destruction
of \( J/\psi  \) and \( \psi ^{\prime } \) needs \( \omega _{\mathrm{min}}=0.7 \)
GeV and 0.1 GeV, respectively. However, as long as the premeson are still in
formation, the uncertainty relation tells us that the binding energies are not
yet the final ones.

The choice of \( \omega _{\mathrm{min}} \) deserves a special discussion since
it is related to the problem of the energy dependence of \( \Psi  \)-\( g \)
break-up cross section \( \sigma _{\mathrm{abs}}^{\Psi g} \) in Eq. (\ref{I1Eq}).
At high energies it is dominated by gluonic exchange in \( t \)-channel and
slightly grows with energy\cite{Hue98a}. The inelastic \( \Psi  \)-\( g \)
cross section is by a color factor 9/4 different from the \( \Psi  \)-\( q \)
cross section, which is according to the constituent quark model related to
the \( \Psi  \)-N cross section. Therefore, one expects
\[
\sigma _{\mathrm{abs}}^{\Psi g}\simeq \frac{9}{4}\sigma _{\mathrm{abs}}^{\Psi q}\simeq \frac{3}{4}\sigma _{\mathrm{abs}}^{\Psi N}\, \, .\]

\begin{figure}
{\par\centering \resizebox*{6cm}{6cm}{\includegraphics{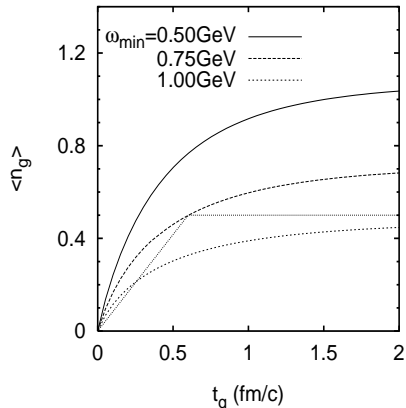}} \par}

\caption{Mean number of prompt gluons as a function of the time \protect\( t_{g}\protect \)
after the NN collision for various values of \protect\( \omega _{\mathrm{min}}\protect \).\label{ngfig}}
\end{figure}

We are left with one adjustable parameter \( \omega _{\mathrm{min}} \) in Eq.
(\ref{ngeq}). In Fig. \ref{ngfig} we show the mean number of gluons, \( \langle n_{g}\rangle  \),
as a function of length \( t_{g} \) for different choices of parameter \( \omega _{\mathrm{min}} \).
In the numerical results presented in the next Section we will not use the form
of Fig. \ref{ngfig} which is generated by the somewhat arbitrary exponential
form in Eq. (\ref{ngeq}), but use a simplified form of \( \langle n_{g}(t_{g})\rangle  \)
in order to reduce the numerical efforts:
\[
\langle n_{g}(t_{g})\rangle =\left\{ \begin{array}{ll}
n_{g}^{0}\, t_{g}/t_{0}\, \, \, \, \, \,  & \mathrm{if}\, \, t_{g}<t_{0}\, ,\\
n_{g}^{0}\, \, \,  & \mathrm{if}\, \, t_{g}\geq t_{0\, }.
\end{array}\right. \]
 This step function type behaviour is also shown in Fig. \ref{ngfig}. In our
calculations we have used a fixed value \( t_{0}=0.6 \) fm/\( c \), and have
varied \( n_{g}^{0} \) within certain limits, \( 0.5\leq n_{g}^{0}\leq 1 \)
for the \( J/\psi  \) (corresponding to \( \omega _{\mathrm{min}} \) between
0.5 and 1 GeV) and \( 1\leq n_{g}^{0}\leq 2 \) for the \( \psi ^{\prime } \)
because of its smaller binding energy.

\section{Results and discussions}

\begin{figure}
{\par\centering \resizebox*{6cm}{6cm}{\includegraphics{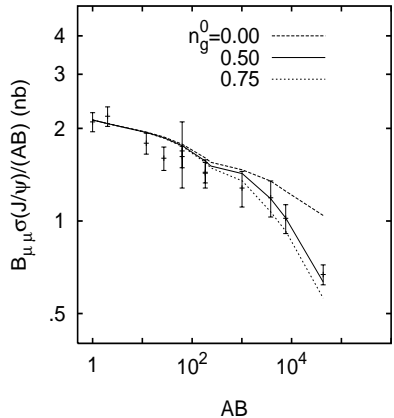}}  \par}

\caption[ABpsifig]{The $ J/\psi$ total cross section as a function of the
product $AB$ of the projectile and target atomic mass numbers
at 200 GeV/c, for $n_{g}^{0}=0,\, 0.5,\, 0.75$. The data
are from \cite{NA5097}\label{ABpsiFig}.
}
\end{figure}

We have calculated the total cross section of \( J/\psi  \) production in proton-nucleus
and nucleus-nucleus reactions according to Eq. (\ref{SABEq}), and the results
are shown in Fig. \ref{ABpsiFig} for three different values of \( n_{g}^{0} \).
The absolute values of cross section are normalized to the mean value of those
in p-p and p-d reactions AB\( =1 \) and 2. The long-dashed curve corresponds
to the result when there are no gluons (only suppression due to nucleons), and
cannot reproduce the experimental data for the Pb-Pb reactions, while the solid
line with \( n_{g}^{0}=0.5 \) accounts for all the data very well. 

\begin{figure}
{\par\centering \resizebox*{6cm}{6cm}{\includegraphics{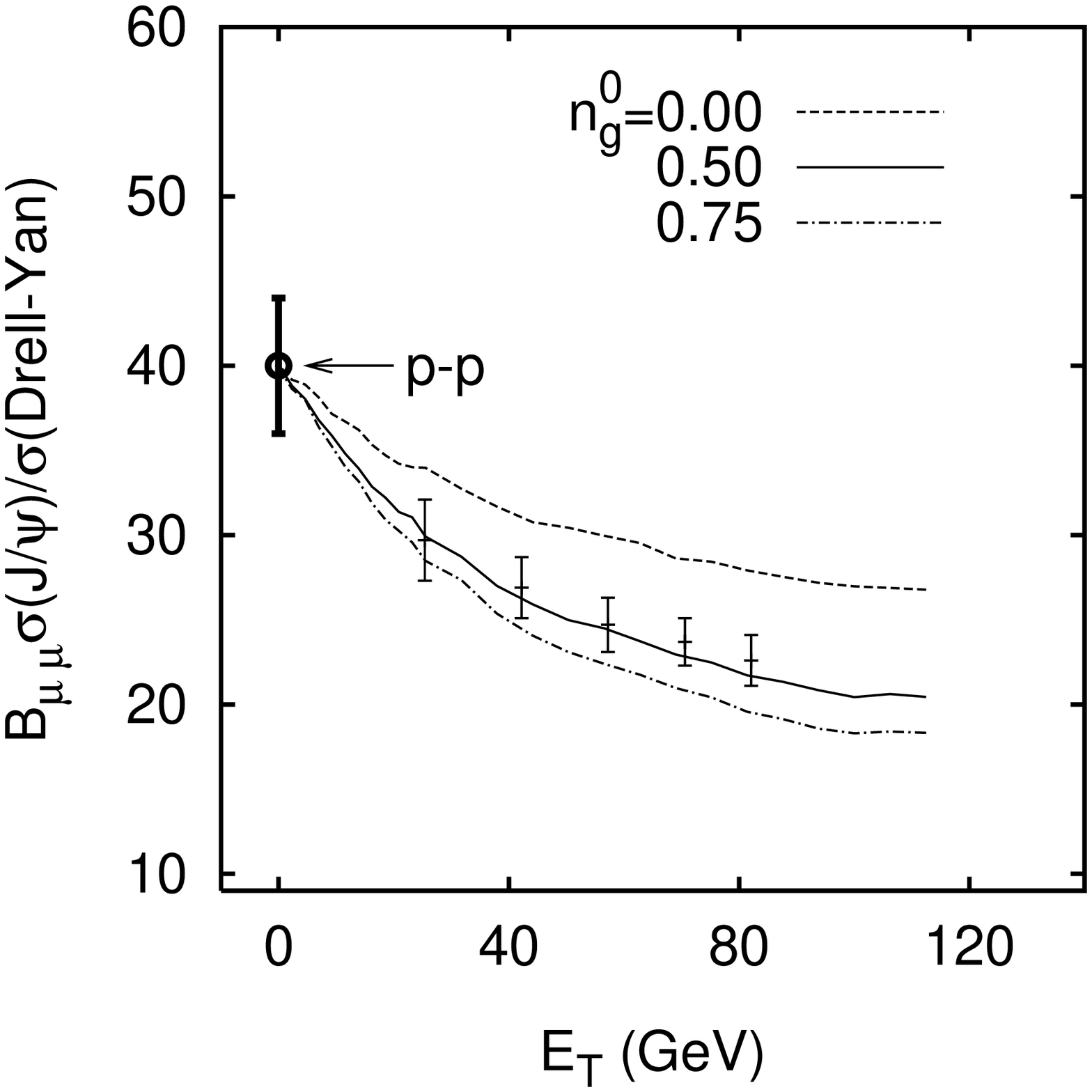}}  \par}

\caption[SUETpsifig]{Ratio of $J/\psi$ to Drell-Yan cross section as a function
of centrality in S-U collisions at 200 GeV/c, with $ n_{g}^{0}=0,\, 0.5,\,
0.75$.
The data points are from \cite{NA3899}.\label{SUETpsiFig} The calculated curves
are normalized at $E_{T}=0$ to the ratio observed for p-p
collisions.}
\end{figure}

Various experiments measure charmonium production as a function of the transverse
energy \( E_{T} \) which corresponds to different centralities. Fig.\ref{SUETpsiFig}
shows the calculated suppression in S-U collisions as a function of \( E_{T} \)
with the normalization at \( E_{T}=0 \) (p-p data). Note that the rather large
uncertainty introduced by this normalization.

\begin{figure}
{\par\centering \resizebox*{6cm}{6cm}{\includegraphics{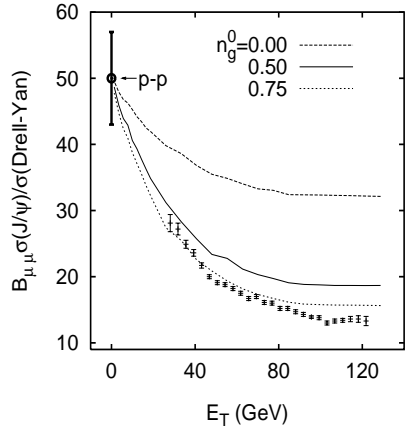}}  \par}

\caption[PbPbETpsifig]{Ratio of \protect\( J/\psi \protect \) to Drell-Yan cross section as a function
of centrality in Pb-Pb collisions at 158 GeV/c for \protect\( n_{g}^{0}=0,\, 0.5,\, 0.75\protect \).
The data are from \cite{NA3899}. The calculated curves are normalized at \protect\( E_{T}=0\protect \)
to the ratio observed for p-p collisions.\label{PbPbETpsiFig}}
\end{figure}

\begin{figure}
{\par\centering \resizebox*{6cm}{6cm}{\includegraphics{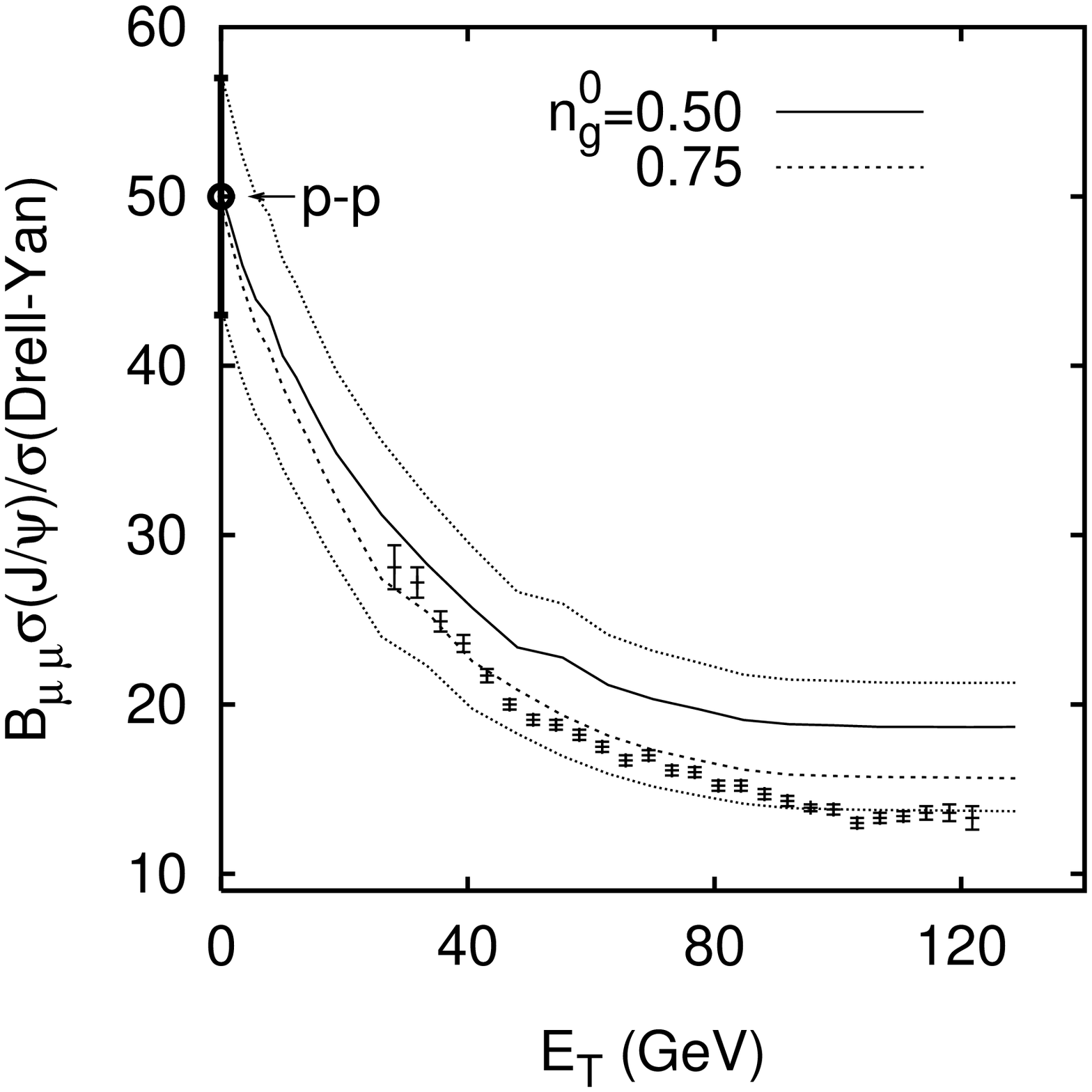}}  \par}

\caption{Ratios of \protect\( J/\psi \protect \) to Drell-Yan cross section as a function
of centrality in Pb-Pb collisions at 158 GeV/c for \protect\( n_{g}^{0}=0.5,\, 0.75\protect \)
(the two curves in the middle). The upper curve corresponds to the result for
\protect\( n_{g}^{0}=0.5\protect \) but normalized at \protect\( E_{T}=0\protect \)
to p-p data point with a standard deviation higher, and the lower one is the
\protect\( n_{g}^{0}=0.75\protect \) curve normalized at \protect\( E_{T}=0\protect \)
to p-p data point with a standard deviation lower.\label{PbPbETpsiNormFig}}
\end{figure}

A comparison of our calculation with NA50 data for Pb-Pb collisions at 158 GeV/\( c \)
is presented in Fig. \ref{PbPbETpsiFig} for \( n_{g}^{0}=0,\, 0.5,\, 0.75 \).
As for S-U the calculation is normalized at \( E_{T}=0 \) to the p-p point.
It is obvious that a theory with no gluons cannot reproduce the data. The value
\( n_{g}^{0}=0.5 \) which is predicted for \( \omega _{\mathrm{min}}=0.75 \)
(which corresponds to the \( J/\psi  \) binding energy) fits the absolute \( J/\psi  \)
production cross sections (Fig. \ref{ABpsiFig}) well, but seems a little high
in the \( E_{T} \) distribution for Pb-Pb collisions. However, one also has
to take into account that there is a nearly 15\% uncertainty in the normalization
(see Fig. \ref{PbPbETpsiNormFig}).

\begin{figure}
{\par\centering \resizebox*{6cm}{6cm}{\includegraphics{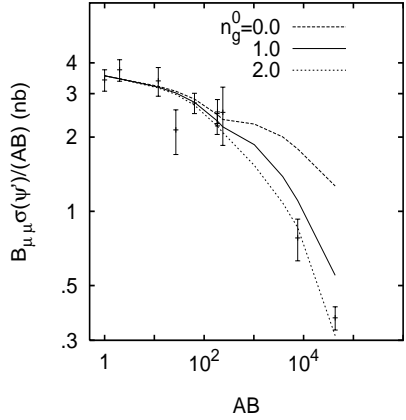}} \par}

\caption{The integrated \protect\( \psi ^{\prime }\protect \) production cross section
as a function of the product \protect\( AB\protect \) of the projectile and
target atomic mass numbers at 200 GeV/c.\label{ABpsipFig}}
\end{figure}

\begin{figure}
{\par\centering \resizebox*{6cm}{6cm}{\includegraphics{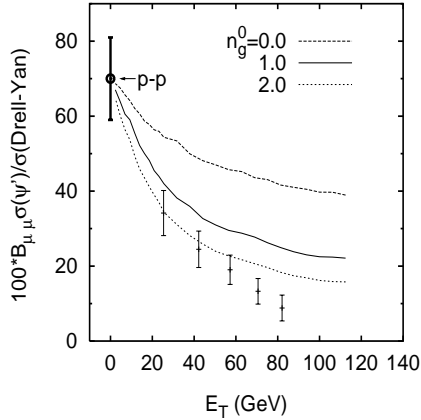}} \par}

\caption[SUETpsipfig]{The \protect\( \psi ^{\prime }\protect \) cross section relative to the Drell-Yan
cross section as a function of the transverse energy for S-U collisions at 200
GeV/c\cite{NA3898}. \label{SUETpsipFig}}
\end{figure}

We proceed to the suppression of \( \psi ^{\prime } \) in nuclear collisions.
Since the \( \psi ^{\prime } \) is less bound than the \( J/\psi  \), we may
take \( \omega _{\mathrm{min}}\geq 200 \) GeV and correspondingly \( n_{g}^{0}\simeq 1 \)-2
is the theoretical choice and leads to reasonable description of the absolute
production cross section (Fig. \ref{ABpsipFig}), but the \( E_{T} \) distribution
in S-U is not well reproduced (Fig.\ref{SUETpsipFig}). The data seem to fall
faster than the calculation. 

\begin{figure}
{\par\centering \resizebox*{6cm}{6cm}{\includegraphics{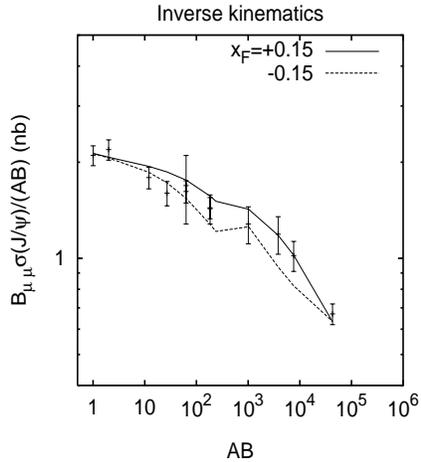}}  \par}

\caption{The integrated \protect\( J/\psi \protect \) total cross sections in nuclear
collisions in the experimental configuration (\protect\( x_{F}=0.15\protect \))
and in the situation of inverse kinematics (\protect\( x_{F}=-0.15\protect \))
.\label{ABpsiInversFig}}
\end{figure}

\begin{figure}
{\par\centering \resizebox*{6cm}{6cm}{\includegraphics{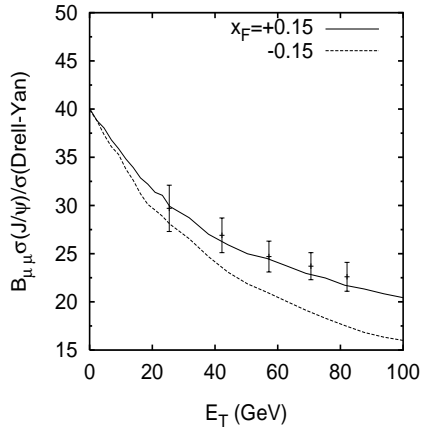}}  \par}

\caption{Transverse energy distribution of \protect\( J/\psi \protect \) production
relative to the Drell-Yan for S-U collisions (\protect\( x_{F}=0.15)\protect \)
and hypothetical U-S collisions (\protect\( x_{F}=-0.15)\protect \).\label{SUETpsiInversFig}}
\end{figure}

Finally, we have calculated the integrated and \( E_{T} \) dependent cross
sections of \( J/\psi  \) production for the inverse kinematics, \emph{i.e.}
p-A and A-p collisions and S-U and U-S collisions. We expect differences, since
the charmonium is observed with a finite rapidity with respect to the NN c.m.
system. In our calculation we just exchange \( v_{cm}\rightarrow -v_{cm} \)
of the charmonium. The differences arise because of the formation and production
times. Figs. \ref{ABpsiInversFig} and \ref{SUETpsiInversFig} show two results.
The effects of inverse kinematics are significant but not dramatic.

\section{Summary and conclusions}

We have calculated charmonium suppression in nuclear collisions for \( J/\psi  \)
and \( \psi ^{\prime } \) and for integrated cross sections and differential
ones with respect to transverse energy. 

The emphasis of the present paper has been on the contribution of the prompt
gluons to the suppression. This contribution turns out to be significant and
decisive, and can quantitatively account for the suppressions in almost all
cases. However, since the values of the input variables \( n_{g}^{0} \) and
\( \sigma ^{\Psi g} \) bear considerable uncertainty, we cannot exclude that
other mechanisms (plasma, hadronic comovers) will also contribute. In these
calculations several parameters are adjusted so that the relative importance
of the various contributions, including those of prompt gluons, cannot be quantitatively
determined at the moment.

\section*{Acknowledgments}

We thank C. Gerschel and A. Mueller for several comments. This paper has been
supported in part by a grant from the BMBF under contract number 06 HD 742.

\end{document}